\documentclass[12pt]{article}

\usepackage{feyn}
\usepackage{amsmath,amsfonts,amssymb,cancel}
\usepackage{graphicx}
\usepackage{booktabs}
\usepackage{bm}
\usepackage{float}
\usepackage{slashed}
\usepackage{subfigure}

\newcommand{\be}{\begin{equation}} 
\newcommand{\ee}{\end{equation}} 
\newcommand{\bea}{\begin{eqnarray}}  
\newcommand{\eea}{\end{eqnarray}}
\newcommand{\bs}{\begin{split}} 
\newcommand{\es}{\end{split}}

\newcommand{\dslash}{\cancel{\partial}~}
\renewcommand{\imath}{i}

\newcommand{\units}[1]{~\mathrm{#1}}

\newcommand{\ctoprule}{\toprule[0.5mm]}
\newcommand{\cbottomrule}{\bottomrule[0.5mm]}
\newcommand{\cmrule}{\midrule[0.25mm]}

\newcommand{\mt}[1]{\mathrm{#1}}

\begin{document}

%----------------------------------- TITLE AND AUTHORS -----------------------------------------%

\begin{center}

\renewcommand{\thefootnote}{\fnsymbol{footnote}}
\setcounter{footnote}{1}

{\Large {\bf LHC Signals of Non-Custodial Warped 5D Models}}\\
\vspace*{1.cm}
{\bf Jorge de Blas}\footnote{jdeblasm@nd.edu},
{\bf Antonio Delgado}\footnote{antonio.delgado@nd.edu},
{\bf Bryan Ostdiek}\footnote{bostdiek@nd.edu} and \\
{\bf Alejandro de la Puente}\footnote{adelapue@nd.edu}

\vspace{0.5cm}

Department of Physics, University of Notre Dame,\\
Notre Dame, IN 46556, USA

\end{center}
\vspace{0.75cm}

\renewcommand{\thefootnote}{\arabic{footnote}}
\setcounter{footnote}{0}

%--------------------------------------------- ABSTRACT ---------------------------------------------%

\begin{abstract}

We study the implications at the LHC for a recent class of non-custodial warped extra-dimensional models where the $AdS_5$ metric is modified near the infrared brane. Such models allow for TeV Kaluza-Klein excitations without conflict with electroweak precision tests. We discuss both the production of electroweak and strong Kaluza-Klein gauge bosons. As we will show, only signals involving the third generation of quarks seem to be feasible in order to probe this scenario. 

\end{abstract}

%-------------------------------- DOCUMENT: INTRODUCTION ---------------------------------%

\section{Introduction}
\label{Intro}

One of the big mysteries in high energy physics is the large hierarchy between the Planck scale and the electroweak (EW) scale. This has been the inspiration for many theories beyond the Standard Model (SM). One intriguing solution is to assume the existence of extra dimensions. In this case the geometry of the dimensions can provide a natural explanation for the hierarchy. The original Randall-Sundrum (RS) scenario \cite{Randall:1999ee} involves one warped extra spatial dimension. The model is built on a slice of $AdS_5$, bounded by two four-dimensional branes, namely the ultraviolet (UV) or Planck brane and the infrared (IR) or TeV brane. The Higgs field is localized on the IR brane so, by adjusting the size of the extra dimension, the warping can redshift the Planck scale to the EW scale. 

The RS model has been extensively studied. The original model had all of the SM particles localized on the IR brane. However, only the Higgs needs to be localized near the IR brane to solve the hierarchy problem. The other particles can live in the bulk. In that case, for each field propagating through the 5D bulk we have an infinite tower of replicas with increasing masses, known as Kaluza-Klein (KK) excitations. However, various problems can arise when placing the other SM particles in the bulk. Placing the gauge bosons in the bulk can lead to parametrically large contributions to the Peskin-Takeuchi \textit{S} and \textit{T} parameters \cite{Peskin:1990zt} coming from the KK excitations \cite{Csaki:2002gy}. This pushes the masses of the KK modes toward large values so the corrections are small. But then the new states would not be detectable at the Large Hadron Collider (LHC), which makes the model less phenomenologically appealing. Moreover, the presence of very large masses would result in a little hierarchy problem. Explaining the stability of the EW scale would reintroduce a certain amount of fine tuning in the theory, going against one of the main motivations of the warped extra dimensional scenario. Another extension is to add the fermions to the bulk. This softens the contributions to the \textit{S} and \textit{T} parameters and provides a mechanism for generating the hierarchy of fermion masses without assuming any fundamental hierarchy in the 5D Yukawa couplings \cite{Gherghetta:2000qt,Grossman:1999ra,Huber:2000ie}.

Further extensions involve enlarging the SM symmetry by gauging custodial symmetry in the bulk. This kind of models allow for lower masses of the KK modes while still protecting the \textit{T} parameter \cite{Agashe:2003zs}. In particular, ref. \cite{Agashe:2003zs}  shows that it is possible to fit the body of electroweak precision data (EWPD) and have KK mode masses around 3 TeV. The authors of \cite{Agashe:2007ki,Agashe:2008jb} looked at the LHC signals for the KK modes of the gauge bosons in this model with custodial symmetry. Their results show that discovering the KK modes is not an easy task and would require high integrated luminosity and $\sqrt{s}=14\units{TeV}$. In particular, they show that KK states of mass 2 (3) TeV can be discovered with $\sim$100 fb$^{-1}$ ($\sim$ 1 ab $^{-1}$) of integrated luminosity for the neutral KK modes, and $\sim$100 (300) fb$^{-1}$ for the charged KK modes.

Although gauging custodial symmetry does protect the EW observables, the extra symmetry would imply the existence of several new particles, and thus a more complex spectrum. A more recent class of model looks for an alternate solution. The ``soft wall'' models \cite{Karch:2006pv} - \cite{Cabrer:2009we}  replace the IR boundary with a smooth spacetime cut-off via a modified warp factor. In this paper we focus on a class of models using a soft-wall metric within the hard-wall scenario \cite{Cabrer:2010si} - \cite{Cabrer:2011mw}. These kind of models also assume a bulk Higgs instead of an IR one.  Both ingredients allow the softening of EWPD bounds and thus lighter KK states than in the standard RS model. This provides a minimal extension of the SM model within a five dimensional warped scenario, with gauge bosons within the LHC reach. In this paper we study which LHC signals would allow testing the existence of the KK excitations of the SM gauge bosons in such scenarios. As we will show, most of the standard channels offer little chances of discovery, even for $\sqrt{s}=14\units{TeV}$, including the diboson channels proposed in  \cite{Agashe:2003zs}. Only those involving the third family of quarks look promising. For the $t\overline{t}$ channel this has been already studied in the literature \cite{Carmona:2011ib}. The model discussed here is slightly different, with a different profile for the 5D fermion masses. As we will show, after cuts this channel could prove the existence of the KK gluons even for $\sqrt{s}=8$ TeV. We also study the signals for the case of the charged EW KK gauge bosons, which could be used as complementary searches to confirm the extra dimensional nature of this scenario.

In the next section we review the details of the extra dimensional model to be studied in this paper, and discuss the benchmark point used in our analyses. In Section \ref{Prod_Neutral} we study the LHC signals for the case of the first KK modes of the neutral gauge bosons. Section \ref{Prod_Charged} presents the corresponding analysis for the case of the charged states. In all cases we comment on the most common channels and give details on those which are more promising for the LHC searches. Finally, we present our conclusions.

%-------------------------------- DOCUMENT: SECTIONS -----------------------------------------%

\section{Review of 5D Warped Physics}
\label{sec_review}

In this section we briefly introduce the extra dimensional model under consideration. For further details we refer to \cite{Cabrer:2011fb}. The starting point is a five dimensional scenario with gravitational background
\be
ds^2 = e^{-2 A(y)} \eta_{\mu\nu}dx^{\mu}dx^{\nu} - dy^2,
\label{metric_1}
\ee
where $\eta_{\mu\nu} = \text{diag}(1,-1,-1,-1)$. In Eq. (\ref{metric_1}) we denote the extra dimension coordinate as `$y$' while `$x^{\mu}$' refer to the standard four dimensions coordinates. The warp factor is determined by the function $A(y)$, which only depends on the position along the fifth dimension. We consider a finite extra dimension, bounded by the UV and IR branes, placed at $y = 0$ and $y = y_1$, respectively. In the standard RS scenario the warp function is given by
\be 
A_{\mt{RS}}(y)=  ky, \label{eqn_met_RS}
\ee
where $k$ is the $AdS_5$ inverse curvature radius. Here we will consider a modified $AdS_5$ (MAdS$_5$) scenario where the RS metric is modified near the IR brane as follows:
\be
A(y) =  ky - \frac{1}{\nu^2} \log\left(1-\frac{y}{y_s}\right). \label{eqn_met_AdSmod}
\ee
In Eq. (\ref{eqn_met_AdSmod}), $\nu$ is a real arbitrary parameter and we have introduced a singularity at $y_s$, at which the warp factor vanishes. The position is chosen such that $y_s>y_1$ so the curvature singularity is always hidden behind the IR brane. Notice that the metric (\ref{eqn_met_AdSmod}) reduces to the RS one in the limits $\nu \rightarrow \infty$ and $y_s \rightarrow \infty$. 

The relevant field content of the theory for our studies is essentially that of the SM. Upon Kaluza-Klein reduction, the effective four-dimensional theory contains an infinite tower of replicas for each of the SM fields, with increasing masses. For the gauge bosons we take,
\be
A_\mu\left(x,y\right)=\frac{1}{\sqrt{y_1}}\sum_{n=0}^\infty f_A^n(y) A_\mu^{(n)}(x),
\ee
where the 5D profiles satisfy the differential equation and Neumann boundary conditions
\be
\partial_y^2 f_A^{n}(y) - 2 \partial_y A(y) \partial_y f_A^{n}(y)+ m_n^2 e^{2A(y)} f^n(y) = 0,~~~\left.\partial_y f_A^n\right|_{y=0,y_1}=0,
\ee
together with the following normalization condition
\be
\frac{1}{y_1}\int_0^{y_1} dy f_A^n f_A^m=\delta_{nm}.
\ee
Instead of placing the Higgs on the IR brane we will consider a delocalized bulk Higgs,
\be
H(x,y)=\frac{1}{\sqrt{2}}e^{i\chi(x,y)}\left(\begin{array}{c} 0 \\ h(y)+\xi(x,y)\end{array}\right).
\label{5DHiggs}
\ee
In Eq. (\ref{5DHiggs}) $\chi(x,y)$ and $\xi(x,y)$ contain the Goldstone bosons and the physical Higgs excitations, respectively, while $h(y)$ denotes the $y$-dependent Higgs background. For a light Higgs mass, most of the vacuum expectation value is carried by the zero mode, which is also approximately proportional to $h(y)$. With a suitable bulk Higgs mass the background is given by
\be
h(y)=c_1e^{aky}\left(1+c_2 \int_0^y dy^\prime e^{4A(y^\prime)-2aky^\prime}\right),
\ee
where the parameter $a$ controls the localization. In order to maintain the exponential behaviour $h(y)\sim e^{aky}$, and thus preserve the RS solution to the hierarchy problem $a\gtrsim a_0=2A(y_1)/ky_1$ \cite{Cabrer:2011fb,Cabrer:2011vu}.

The constraints from EW precision observables are dominated by those from oblique parameters, and, in particular, for masses $M_{KK}\gtrsim 1\units{TeV}$ the ones from the $S$ and $T$ parameters. When the EW gauge bosons propagate in the bulk, the tree-level contributions from the KK states to these parameters read \cite{Cabrer:2011fb},
\be
\alpha S= 8 s_W^2 c_W^2 M_Z^2\frac{I_1}{\rho^2}\frac{1}{Z},~~~~~~~
\alpha T= s_W^2 M_Z^2 \frac{I_2}{\rho^2}\frac{k y_1}{Z^2},~~~~~~~
W=Y=c_W^2 M_Z^2\frac{I_0}{\rho^2}\frac{1}{k y_1},
\label{STWY}
\ee
where $\rho\equiv k e^{-A(y_1)}$ is the IR scale, and we have introduced
$$Z=k\int_0^{y_1}dy \frac{h^2(y)}{h^2(y_1)}e^{-2A(y)+2A(y_1)},$$
as well as the dimensionless integrals
$$I_n=k^3\int_0^{y_1}(y_1-y)^{2-n} u^n(y) e^{2A(y)-2A(y_1)},~~~n=0,1,2,$$
with $u(y)=\int_y^{y_1}dy^\prime \frac{h^2(y^\prime)}{h^2(y_1)}e^{-2A(y^\prime)+2A(y_1)}$.

For a light Higgs the quantity $\sqrt{Z}$ renormalizes the wave function of the Higgs in the effective 4D Lagrangian. In particular, the coupling of the gauge KK modes to the Higgs current is proportional to $Z^{-1}$, which explains the different dependence on $Z$ for the $S$, $T$, $W$ and $Y$ parameters.

As can be deduced from Eq. (\ref{STWY}) the most stringent constraints will come from $T$, which is volume enhanced, and $S$, which is volume independent. Contributions to $W$ and $Y$, on the other hand, are volume suppressed. In this regard, notice that with the modified metric $ky_1<A(y_1)\sim 35$, i.e.,  we can still solve the hierarchy with a reduced volume of the extra dimension. This softens a little bit the enhancement of the $T$ parameter. The largest suppression, however, comes from the wave function renormalization, as it turns out that with the modified metric $Z$ can have large values for large regions in the parameter space \cite{Cabrer:2011fb}. As will be shown below, this suppression can be traced back to the fact that, with the new metric, while the KK gauge bosons are shifted to the IR, the Higgs is pushed into the bulk. This naturally suppresses the mixing between the SM gauge fields (zero modes) and the KK gauge bosons, thus reducing the contributions to the $T$ parameter.

For fermions $\psi=\left(\psi_L, \psi_R\right)^T$, the free part of the action is given by
\be
S_\psi\!=\!\int\! d^5x~\!\left\{e^{-3A} \overline{\psi} i\dslash \psi+e^{-4A}\!\left(\overline{\psi} \gamma_5\partial_y \psi -2 \partial_y A \overline{\psi}\gamma_5 \psi - M_\psi (y) \overline{\psi} \psi\right)\!\right\}.
\ee
Thus, the KK expansion for $\hat \psi_{L,R}=e^{-2A(y)}\psi_{L,R}$,
\be
\hat \psi_{L,R}(x,y)=\frac{1}{\sqrt{y_1}}\sum_{n=0}^\infty f_{L,R}^{n} \hat \psi_{L,R}^{(n)}(x),
\ee
has modes satisfying
\be
\begin{split}
 \partial_y f_L^{n}(y) + M_\psi(y)f_L^{n}(y) &= e^{A(y)} m_n f_R^{n}(y),
\\
 \partial_y f_R^{n}(y) - M_\psi(y) f_R^{n}(y) &= -e^{A(y)} m_n f_L^{n}(y),
 \\
\frac{1}{y_1} \int_0^{y_1}dy f_{L,R}^n f_{L,R}^m e^{A(y)}&= \delta_{nm}.
\end{split}
\label{EqFerm}
\ee
As the 5D Dirac equation couples both chiralities, only the boundary condition for one of the components must be specified. We impose Dirichlet boundary conditions in both branes over the corresponding chirality in a way that the zero modes reproduce the chiral matter content of the SM.

Following \cite{Cabrer:2011qb} we choose the $y$-dependent bulk mass to be $M_\psi(y)=\pm c_\psi A'(y)$, where the ``$+$''(``$-$'') sign is for left-handed (right-handed) zero modes. In this way, the properly normalized fermion zero modes (see Eq. (\ref{EqFerm})),
\be
{\tilde f}_{L,R}^{0}\equiv e^{\frac{A(y)}{2}}f_{L,R}^{0}=\frac{e^{(\frac 12 - c_\psi) A(y)}}{ \left[\frac{1}{y_1}\int_0^{y_1}e^{A(y^\prime)(1- 2c_\psi)}dy^\prime\right]^{1/2}},
\ee
are localized to the IR (UV) for values of $c_\psi<1/2$ ($c_\psi>1/2$).\footnote{This choice can be compared to \cite{Carmona:2011ib} where the authors chose a RS-like profile for the bulk mass, $M_\psi=c_\psi k$. For that choice one can still control the localization of fermions by choosing the adequate $c$ values. But, for instance, one cannot obtain an exactly flat fermionic profile.} By choosing the adequate localization for the different fermion families one can obtain the pattern of SM Yukawa couplings in the effective 4D theory without introducing any hierarchy in the fundamental theory. It turns out that, compared to the RS case, fitting the fermion masses and mixings yields less IR-localized fermion profiles \cite{Cabrer:2011qb}, which in turn reduces the couplings of fermion zero modes to the KK gauge bosons. On the other hand, using different localizations usually generates contributions to non-oblique observables, and it can also generate flavour changing neutral currents (FCNC) and extra sources of CP violation. These pose extra constraints on the scale at which the extra dimensional physics can manifest and, in particular, imply stronger bounds than those from oblique observables. Still, as a result of the preference for less IR-localized fermions, these bounds are milder than in the RS scenario.

\subsection{Benchmark points}

As explained in \cite{Cabrer:2011fb}, when choosing the values of the model parameters it is convenient to trade $y_s$ for the value of the 5D curvature radius at the IR brane in units of $k$: $kL_1$. This parameterization makes easier to keep track of the dependence of the tree-level corrections to EW precision observables on the model parameters. In particular, such corrections are reduced for lower values of $a$, $\nu$ and $kL_1$ (as they yield larger values for $Z$). On the other hand, $kL_1$ is bounded from below, $kL_1\gtrsim 0.2$, if we want to require perturbativity of the gravitational expansion, as otherwise this would introduce a large backreaction \cite{Cabrer:2011fb}. The free parameters in the metric are also constrained to reproduce the adequate hierarchy between the UV and IR branes, i.e., $A(y_1)\approx 35$. This condition can be used to fix the IR brane position, $y_1$, given values for the other parameters. Finally, as explained above, $a$ is subject to the constraint $a\gtrsim a_0=2A(y_1)/ky_1$.

We use the benchmark points from Table 1 in \cite{Cabrer:2011fb}. These and the value of $a$ are shown for illustrational purposes in Table \ref{tab_SWparameters}. We also show the typical KK mass scale for the gauge sector, $M_{KK}$. This gives the masses of the first KK modes for the gauge bosons, up to EW symmetry breaking effects. For instance, for the first benchmark point ---which we will use for our phenomenological analyses--- the exact values of the masses for the first KK modes of the gauge bosons are
\be
M_{A_{KK},G_{KK}}=2442\units{GeV},~~M_{Z_{KK}}=2466\units{GeV}~~\mt{and}~~M_{W_{KK}^\pm}=2461\units{GeV}.
\ee
\begin{table}[h]
\centering
ÊÊÊÊ\begin{tabular}{ c  c  cc  cc  cc cc c  c  }
	\ctoprule
ÊÊÊÊÊÊÊ	\!\!BenchMark\!\! & &	$kL_1$		&&  $ky_1$		&& $(ky_s)$		&&  $\nu$	&& $a$ &$M_{KK}$[TeV]\\ 
	\cmrule
ÊÊÊÊÊÊÊÊ1			&&	$0.3$	&& $25$	&& $(26.3)$	&& 0.55	&& 2.8 & 2.4 \\ 
	2			&&	$0.4$ 	&& $28$	&& $(29.6)$	&& 0.64	&& 2.5 & 4.0 \\ 
	3			&&	$0.5$	&& $30$	&& $(31.7)$	&& 0.73	&& 2.4 & 5.2 \\
	 \cbottomrule
ÊÊÊÊ\end{tabular}
\caption{Values of different relevant model parameters and the corresponding KK mass scale for the first KK mode of the gauge bosons, $M_{KK}$.\label{tab_SWparameters}}
\end{table}

\begin{figure}[ht]
\begin{center}
\includegraphics[width=.7\columnwidth]{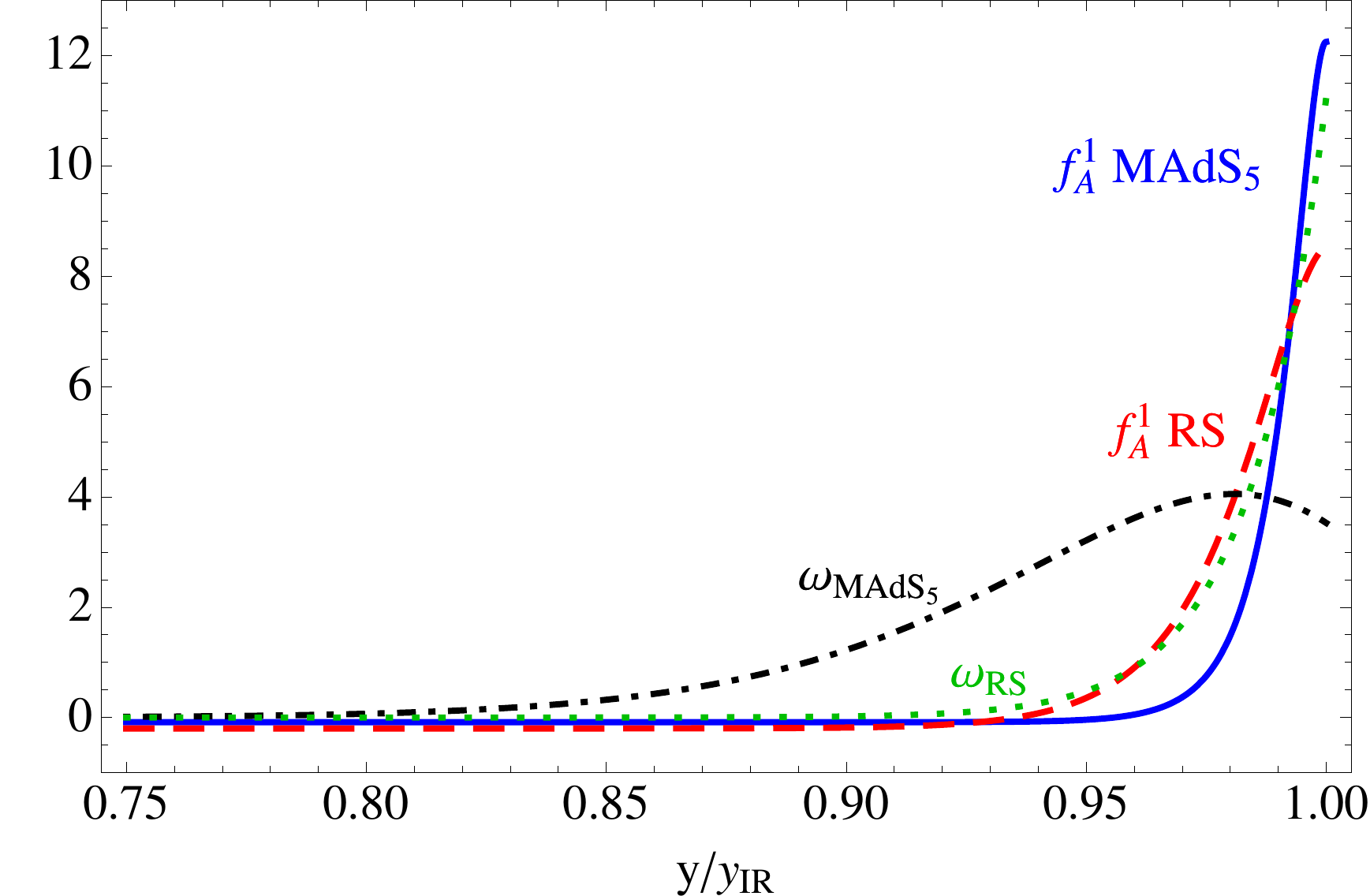}
\caption{Comparison of the first photon KK modes in the RS and MAdS$_5$ scenarios near the IR (red-dashed and blue-solid lines, respectively). We also compare the Higgs wavefunctions in both scenarios for $a=2.8$ (green-dotted and black dot-dashed lines, for the RS and MAdS$_5$ profiles, respectively).}
\label{fig_1stKKMode}
\end{center}
\end{figure}

The effect of the singularity pushes the gauge bosons profiles closer to the IR brane, compared to the RS scenario. In Figure \ref{fig_1stKKMode} we compare, for instance, the profiles for the first KK mode of the photon in the first benchmark point. Even though the profiles show similar structure, with the warp factor in Eq. (\ref{eqn_met_AdSmod}) the profile is more localized towards the IR brane. This makes the interaction between the KK modes and the SM particles weaker, unless the latter are also strongly localized to the IR. On the other hand, as explained above and can also be seen in Figure \ref{fig_1stKKMode}, for a bulk Higgs the singularity ``decouples'' the Higgs background and wavefunction $\omega(y)\sim e^{-A(y)}h(y)$ from the IR. This reduces the overlapping with the KK gauge boson profiles. In particular, for the case of the $W^\pm$ and $Z$ KK modes this helps to protect the contributions to the $T$ parameter without resorting to gauging custodial isospin.

\begin{figure}[ht]
\centering
\subfigure[]{
\includegraphics[width=.478 \textwidth,]{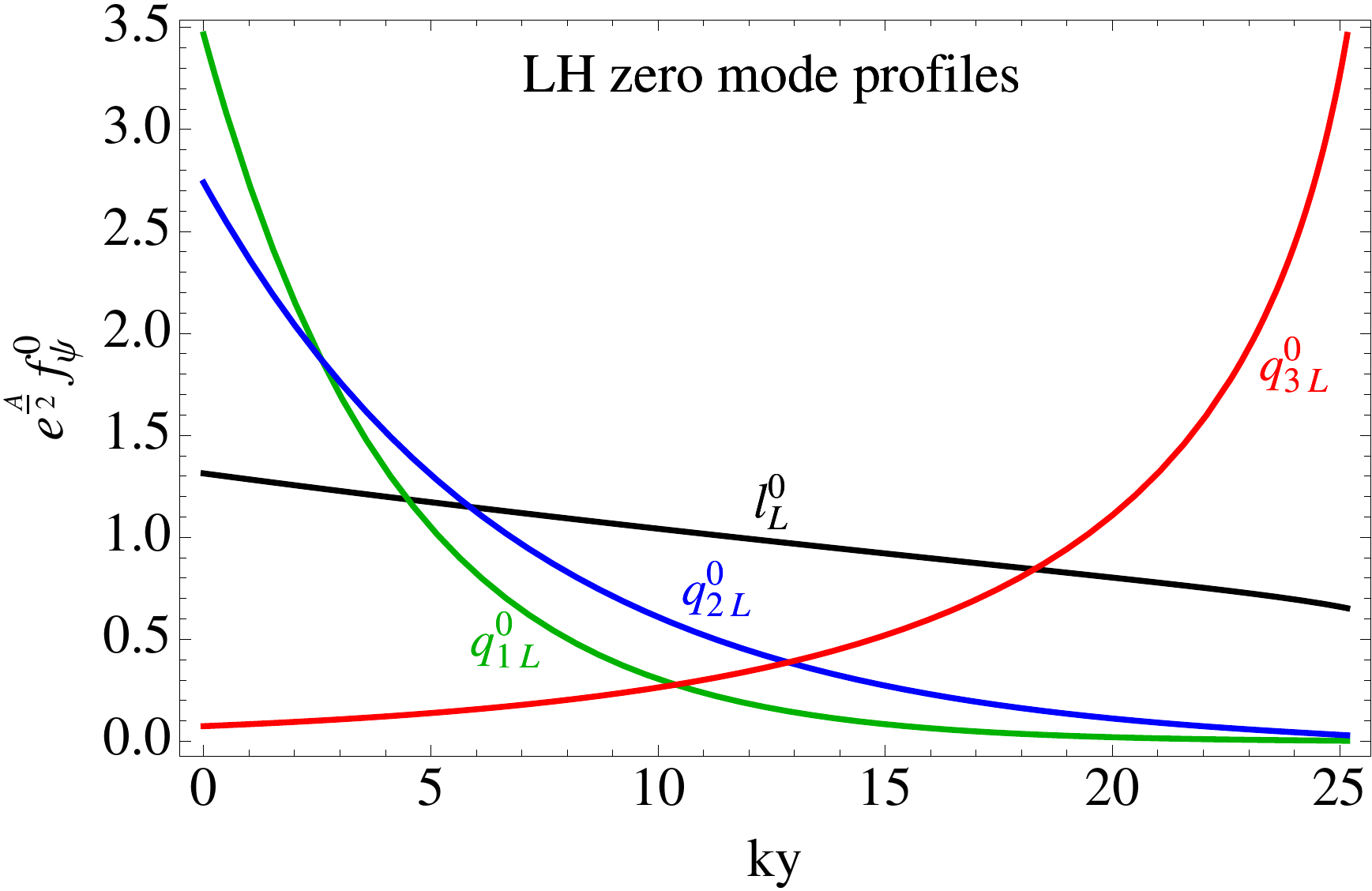}
\label{fig_LHprof}
}
\subfigure[]{
\includegraphics[width=.478 \textwidth]{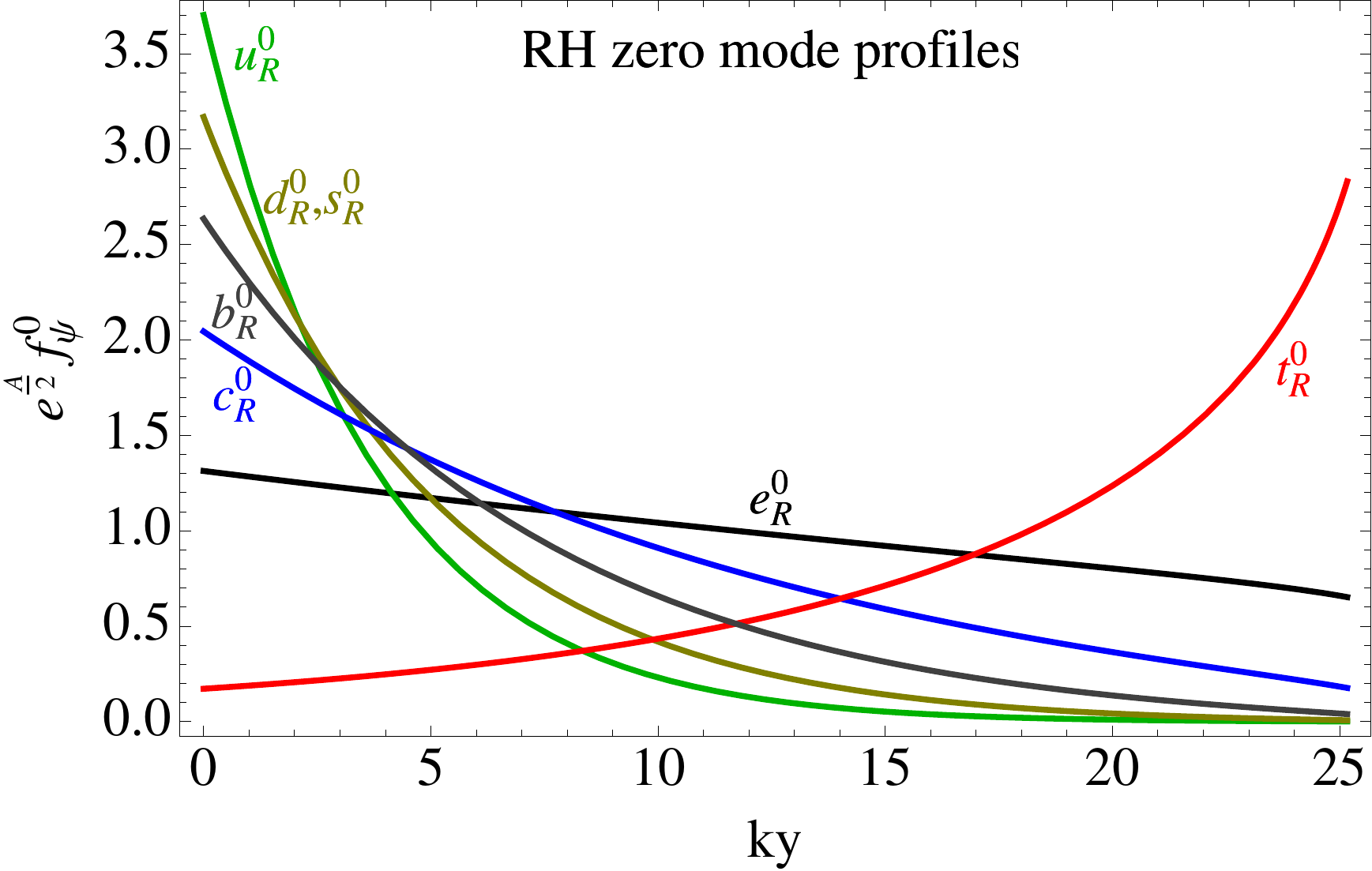}
\label{fig_RHprof}
}
\caption[]{Fermionic profiles for the set of c-values chosen in our benchmark point. \subref{fig_LHprof} Left-handed zero modes. \subref{fig_RHprof} Right-handed zero modes.}
\label{fig_Fermion0Modes}
\end{figure}

Regarding the fermion localizations we follow the results of \cite{Cabrer:2011qb}, where the authors randomly picked 5D quark Yukawa couplings and then fit to the corresponding 4D Yukawa couplings varying the c-values for a model with $\nu = 0.5$ and $k(y_s-y_1)=1$. For the light families we take the central values obtained from their fit. For the third family we use a slightly larger value for $c_{(t,b)_L}$, in order to remain consistent with the $Zb\overline{b}$ constraints for our value of the KK scale. From Figure 4 right in \cite{Cabrer:2011qb}, we choose to increase its value to 0.39. This is still within 1$~\!\sigma$ of the best fit value. As can be deduced from that figure, this also requires readjusting $c_{b_R}$ to reproduce the bottom mass with a 5D Yukawa $\sqrt{k} Y_b^{5D}\sim 1$.\footnote{We do not readjust the c-value for the $t_R$ but instead use a shift in $\sqrt{k} Y_t^{5D}$ (still $\sim {\cal O}(1)$) to reproduce the top mass.} In order to satisfy this constraint we choose a value of $c_{b_R}=0.62$ (1.3$~\!\sigma$ away from the best fit value). The explicit c-values for the quarks are given in Table \ref{tab_quark_cValues}. For the leptons we choose to keep all of them in the same localization to prevent contributions to lepton flavour violating processes. We have taken $c_{(\ell,\nu_\ell)_L}=c_{\ell_R}=0.52$, $\ell=e,\mu,\tau$, which corresponds to nearly flat lepton profiles. This choice minimizes the overlapping between the fermion zero modes and the KK gauge boson profiles. Hence, it also minimizes the contributions to leptonic EW precision observables from four-fermion operators. Of course, this will in turn reduce the phenomenology of the model in the leptonic channels. This is however a characteristic feature of warped extra dimensions where leptons are usually localized towards the UV and predicts small leptonic couplings. The different localizations of the fermion zero modes for this set of c-values is illustrated in Figure \ref{fig_Fermion0Modes}.
\begin{table}[h]
\centering
ÊÊÊÊ\begin{tabular}{ c c c }
	\ctoprule
ÊÊÊÊÊÊÊ	$c_{(u,d)_L} = 0.71$	&	$c_{(c,s)_L}= 0.63$	&	$c_{(t,b)_L} = 0.39$ \\
	$c_{u_R} = 0.74$	&	$c_{c_R} = 0.57$	&	$c_{t_R} = 0.42$ \\
	$c_{d_R} = 0.68$	&	$c_{s_R} = 0.67$	&	$c_{b_R} = 0.62 $\\ 
	\cbottomrule
	\end{tabular}
\caption{c-values used for the quarks.\label{tab_quark_cValues}}
\end{table}

In the subsequent sections we will discuss the collider phenomenology of the scenario described here. For the computation of the decay widths and cross sections of the extra gauge bosons, we implemented the model using FeynRules \cite{Christensen:2008py}. We use MadGraph 5 \cite{Alwall:2011uj} to calculate the corresponding cross section and decay widths. The cross section calculations use CTEQ6L1 for the parton distribution functions \cite{Pumplin:2002vw}. In all cases we will restrict our studies to a parton-level analysis.

%--------------------------------------------------------------------------------------------------------------------------------------
%  Neutral gauge bosons
%--------------------------------------------------------------------------------------------------------------------------------------

\section{Neutral Kaluza-Klein gauge bosons}
\label{Prod_Neutral}
\vspace{-0.22cm}

For the values of the model parameters described in the previous section we have first calculated the production cross section of $pp \rightarrow A_{KK},Z_{KK}$. Table \ref{tab_Prod_14TeV} shows the cross section for producing the KK modes at 14 TeV. We will restrict to this case for most of the paper, since the cross sections at $\sqrt{s} = 8$ TeV are too small. 
\begin{table}[h]
\centering
\begin{tabular}{ c c  c c  c c c  }
	\ctoprule
	$M_{KK}$	&~~~&$2.4$ TeV	&~~~& 	$4.0$ TeV	&~~~&	$5.2$ TeV \\
	\cmrule
	$\sigma(pp\rightarrow A_{KK})$					&~~~& 2.98 fb				&~~~& 131 ab			&~~~&	15.0 ab \\
%	\hline
	$\sigma(pp\rightarrow Z_{KK})$					&~~~&  3.72 fb				&~~~& 153 ab			&~~~& 16.1 ab \\
	\cbottomrule
\end{tabular}
\caption{Production cross section at 14 TeV for the different points in Table \ref{tab_SWparameters}.\label{tab_Prod_14TeV}}
\end{table}
In Table \ref{tab_Prod_14TeV} we also show the results for the benchmark points other than the first one for comparison but, for obvious reasons, we only study the signals for $M_{KK} = 2.4$ TeV. Of course, one could look for configurations with even lower KK masses. In this regard, it must be stressed that, according to the results of \cite{Cabrer:2011qb}, $M_{KK} = 2.4$ TeV already corresponds to relatively small regions in the parameter space. Thus, even though some configurations may allow for lighter KK masses, we expect those solutions to be more fine tuned.

For the model parameters in the first benchmark point we have calculated the decay widths of the first KK modes of the photon and the $Z$: $A_{KK}$ and $Z_{KK}$, respectively. We consider decays into $q\overline{q}$, $l\overline{l}$, and $W^+ W^-$. For the case of the $Z_{KK}$, which also couples to the Higgs field, we also considered $Z_{KK} \rightarrow Z \text{ }h$. Regarding the $\psi\overline{\psi}$ decay channels, only decays into the zero modes are kinematically accessible. We have checked that the first KK states for the bulk fermions are heavy enough so there is no phase space for $\psi^0 \psi^1$ decays.  Notice that, when decaying into a gauge boson, the particle will be extremely boosted due to the mass of the KK modes. Thus the outgoing $W$'s or $Z$'s will be mostly longitudinally polarized. The partial decay widths and branching ratios are given in Tables \ref{tab_akk_width} and \ref{tab_zkk_width}. There are many different decay channels for the KK modes but these are clearly dominated by the third family quarks and, to a less extent, decays into gauge bosons.  The total decay widths also turn out to be relatively large compared to the mass splitting between the different KK gauge bosons. Thus, for a given final state the signal for the different KK modes contributing to that channel cannot be distinguished. In particular, any colored final state will be dominated by the KK gluon contribution. We will comment on this below, when discussing hadronic final states.

\begin{table}[h]
\centering
	\begin{tabular}{lc| c c}
		\ctoprule
		Decay	&&Width&	Branching\\
		channel    	&&[GeV]&Ratio	\\
		\cmrule
		$t\overline{t}$	&& 9.938 &	0.563\\
		$W^+W^-$	&& 5.453	&	0.309\\
		$b\overline{b}$&& 1.751	&	0.099\\
		$u\overline{u}$&& 0.157	&	0.009\\
		$c\overline{c}$&& 0.152	&	0.009\\
		$d\overline{d}$&& 0.039 	&	0.002\\
		$s\overline{s}$&& 0.039 	&	0.002\\
		$e^+e^-$		&& 0.037	&	0.002\\
		$\mu^+\mu^-$	&& 0.037	&	0.002\\
		$\tau^+\tau^-$	&& 0.037	&	0.002\\
		\cmrule
		Total			&& 17.641&	\\
		\cbottomrule
	\end{tabular}
\caption{Decays of $A_{KK}$ for $M_{A_{KK}}=2.4\units{TeV}$.\label{tab_akk_width}}
\end{table}
\begin{table}[h]
\centering
	\begin{tabular}{lc| c c}
		\ctoprule
		Decay	&&Width&	Branching\\
	          channel	&&[GeV]&	Ratio\\
		\cmrule
		$b\overline{b}$	&& 15.960                            &	0.493\\
		$t\overline{t}$&& 11.460	                            &	0.354\\
		$W^+W^-$	&& \phantom{0}2.337	&	0.072\\
		$Z$ $h$		&&\phantom{0}1.847	&	0.057\\
		$d\overline{d}$&& \phantom{0}0.179	&	0.006\\
		$s\overline{s}$&& \phantom{0}0.179	&	0.006\\
		$u\overline{u}$	&& \phantom{0}0.140	&	0.004\\
		$c\overline{c}$	&& \phantom{0}0.138	&	0.004\\
		$\nu\overline{\nu}$&&\phantom{0}0.074   &      0.002\\
		$e^+e^-$		&& \phantom{0}0.012	&	~~$4\cdot 10^{-4}$\\
		$\mu^+\mu^-$	&& \phantom{0}0.012 	&	~~$4\cdot 10^{-4}$\\
		$\tau^+\tau^-$	&& \phantom{0}0.012 	&	~~$4\cdot 10^{-4}$\\
		\cmrule
		Total			&& 32.352&	\\
		\cbottomrule
	\end{tabular}
\caption{Decays of the $Z_{KK}$ for $M_{Z_{KK}}=2.4\units{TeV}$.\label{tab_zkk_width}}
\end{table}

The cleanest signals at the LHC come from decays to leptons, so we first discuss the usual dilepton channel $pp\rightarrow \ell^+ \ell^- $. Next we analyze diboson production $pp\rightarrow W^+W^-$. As we will see, as opposed to the custodial models studied in \cite{Agashe:2007ki}, the particular features of the model considered here does not allow testing the presence of the new neutral gauge bosons in these channels even for the LHC at $\sqrt{s} = 14$ TeV. We finally move to the study of hadronic final states, and in particular $t \bar{t}$ production, where there are actual prospects for observing the extra gauge bosons.

\subsection{Dilepton final states: $p~\!p \rightarrow \left\{A_{KK},~\!Z_{KK}\right\} \rightarrow \ell^+~\!\ell^-$ ($\ell=e,\mu$)}

The dilepton channel is one of the golden channels studied when searching for non-leptophobic extra neutral gauge bosons. However, this is not usually the best channel in order to test warped extra dimensional scenarios. The reason is that the lower KK gauge states tend to be more localized near the IR, while the SM leptons (leptonic zero modes) are typically localized towards the UV. Hence, one expects relatively small couplings to leptons which diminish the signal in the dilepton channel. This is even more pronounced in our case, as the effect of the deformation of the metric pushes the profile for the KK gauge bosons even more to the IR compared to the RS scenario, as shown in Figure \ref{fig_1stKKMode}, yielding very small leptonic branching fractions.\footnote{Notice we have also chosen nearly flat leptonic profiles minimizing the resulting contributions to four-fermion interactions, which in particular also suppresses dilepton production. At any rate, the prospects for discovery in this channel are unchanged for more UV-localized leptons. For instance, for $c_{(\ell,\nu_\ell)_L}=c_{\ell_R}=0.7$ the leptonic branching fractions are 0.013 and 0.002 for the $A_{KK}$ and $Z_{KK}$, respectively. This gives a cross section $\sim 3$ times larger, which is still too small.} We can estimate the corresponding cross sections in the narrow width approximation.\footnote{As we will see, for all the first KK modes the total widths lie in the range of $\sim 20-70\units{GeV}$. Thus, we expect the error in the narrow-width estimations to be ${\cal O}(\Gamma_{KK}/M_{KK})\sim 1-3\%$.} These are given in Table \ref{tab_DileptonCS} for the LHC at $\sqrt{s}=14$ TeV.
Even considering a cut in the dilepton invariant mass of 2000 GeV, the resulting numbers are much smaller than the corresponding SM background: $\sigma_{\mt{SM}}(pp\rightarrow \ell^+ \ell^-)_{M_{\ell\ell}\ge2000\units{GeV}}=390 \units{ab}$.

\begin{table}[h]
\centering
ÊÊÊÊ\begin{tabular}{cc}
	\ctoprule
ÊÊÊÊÊÊÊ	$X$&  $\sigma (p p  \rightarrow  X \rightarrow  \ell^+ \ell^-) $[pb] 	\\ 
	\cmrule
ÊÊÊÊÊÊÊÊ	$A_{KK}$	&  $1.26\cdot 10^{-5}$ \\ 
	$Z_{KK}$	&  $2.86\cdot 10^{-6}$ \\ 
	\cmrule
	Both & $1.55 \cdot 10^{-5}$ \\ 
	\cbottomrule
ÊÊÊÊ\end{tabular}
\caption{Cross section for $p p \rightarrow \left\{A_{KK},~\!Z_{KK}\right\}\rightarrow \ell^+ \ell^-$, $\ell=e,\mu$.\label{tab_DileptonCS}}
\end{table}

\subsection{Diboson searches: $p~\!p\rightarrow \left\{A_{KK},~\!Z_{KK}\right\} \rightarrow W^+~\!W^-$}

Unlike the RS scenario with custodial symmetry, in the present model diboson production will not offer good prospects for the discovery of the new neutral gauge bosons. As can be seen from Tables \ref{tab_akk_width} and \ref{tab_zkk_width}, the branching ratio to $W^+W^-$ is small for both the $A_{KK}$ and the $Z_{KK}$. This can be understood from the fact that decays into longitudinally polarized gauge bosons are controlled by the same couplings of the new states to the physical Higgs, which are precisely the ones suppressed by construction in order to soften the bounds from the oblique observables.  (Likewise, the decay channel $Z_{KK}\rightarrow Z h$ will be also suppressed.) The $W^+ W^-$ are produced from the decay of the extra neutral gauge bosons with a cross section $\sigma(pp\rightarrow \left\{A_{KK},Z_{KK}\right\} \rightarrow W^+W^-)=1.19\units{fb}$. The production is then followed by an hadronic or leptonic decay of the $W^\pm$. The corresponding cross sections for the fully leptonic and the semileptonic ---with one of the $W$'s decaying into jets--- channels are shown in Table \ref{tab_KKDibosonCS}. These are the expected cross sections before making any cuts. The semileptonic cross section is around a factor of six larger than the fully-leptonic one. On the other hand, the latter provides a much cleaner signal. At any rate, both cross sections are quite small, and several orders of magnitude below the corresponding SM backgrounds. This would require making cuts in order to increase the significance, so the total cross sections would be even smaller. Those cuts typically reduce the signal cross section to the level of a few ab, which is too small to take into consideration.

\begin{table}[h]
\centering
ÊÊÊÊ\begin{tabular}{cccccc}
	\ctoprule
ÊÊÊÊÊÊÊ	\!\!$X$\!\!&  $\sigma (p p\!  \rightarrow\!  X \! \rightarrow\!  W^+ W^-) $[pb] 	& BR$(W  \!\rightarrow\!  \ell$ $\nu_\ell)$ 	&&Total [pb]\\ 
	\cmrule
ÊÊÊÊÊÊÊÊ	\!\!$A_{KK}$\!\!	&  $ 9.22\cdot 10^{-4}$ 				& 0.216 			&& $4.30\cdot 10^{-5}$ \\ 
	\!\!$Z_{KK}$\!\!	& $ 2.69\cdot 10^{-4}$ 				& 0.216 			&& $1.25\cdot 10^{-5}$ \\ 
	\cmrule
	\!\!Both\!\! & $1.19\cdot 10^{-3}$ 				& - 				&& $5.56 \cdot 10^{-5}$ \\ 
	\cbottomrule
	\\ \ctoprule
ÊÊÊÊÊÊÊ	\!\!$X$\!\!	&  $\sigma (p p\!  \rightarrow\!  X \! \rightarrow\!  W^+ W^-) $[pb] 	& BR$(W  \!\rightarrow\!  \ell$ $\nu_\ell)$  	& BR$(W\!\rightarrow\! j j)$	&Total [pb]\\ 
	\cmrule
ÊÊÊÊÊÊÊÊ	\!\!$A_{KK}$\!\!	& $ 9.22\cdot 10^{-3}$ 						& 0.216 						& 0.676					& $2.69\cdot 10^{-4}$ \\ 
	\!\!$Z_{KK}$\!\! 	& $ 2.69\cdot 10^{-3}$ 						& 0.216 						& 0.676 					& $7.85\cdot 10^{-5}$ \\ 
	\cmrule
	\!\!Both\!\!& $1.19\cdot 10^{-3}$ 						& - 							& -								& $3.48\cdot 10^{-4}$ \\ 
	\cbottomrule
ÊÊÊÊ\end{tabular}
\caption{Cross sections for $p p \rightarrow \left\{A_{KK},~\!Z_{KK}\right\}\rightarrow W^+ W^-$ in the fully-leptonic and semileptonic decay channels.\label{tab_KKDibosonCS}}
\end{table}

\subsection{Hadronic final states}
\label{Prod_Color}

As stressed above, the small cross sections in the dilepton and $W^+W^-$ channels come from the small branching ratios. These in turn are smaller than in the RS case due to the shift towards the IR of the KK profiles. As a result of this shift, the decay widths are dominated by the decays into the third family of quarks, whose profiles are also localized near the IR.

When considering decays into hadronic states, and in particular into the third generation, we must consider not only signals from the production of the KK photons and $Z$'s, but also KK gluons. Actually, these provide the leading new physics contributions. Since both the photons and gluons are massless, their KK states share the same masses and 5D profiles. The different KK gluon decay channels as well as the corresponding widths and branching ratios are shown in Table \ref{tab_GKKWidth}. As in the case of the $A_{KK}$ and $Z_{KK}$, the stronger localization of this mode to the IR compared to the RS case makes it only worth looking for signals involving the third generation of quarks. Indeed, since light quarks are localized towards the UV, similarly to what happened in the discussion of dilepton final states, the (much less clean) dijet channel will be largely suppressed, which renders it useless for the KK gauge boson searches.
\begin{table}[h]
\centering
ÊÊÊÊ\begin{tabular}{l | c c }
	\ctoprule
ÊÊÊÊÊÊÊ	Decay		&	Width	&	Branching\\ 
ÊÊÊÊÊÊÊ	channel		&	[GeV]         	&	Ratio \\ 
	\cmrule
	$t \overline{t}$	&	40.72	&	0.566	\\
	$b \overline{b}$&	28.70	&	0.399	\\
	$u \overline{u}$&	\phantom{0}0.645	&	0.009	\\
	$d \overline{d}$&	\phantom{0}0.645	&	0.009	\\
	$s \overline{s}$	&	\phantom{0}0.644	&	0.009	\\
	$c \overline{c}$	&	\phantom{0}0.622	&	0.009	\\
	\cmrule
	Total			&	71.979	&		\\
	\cbottomrule
	\end{tabular}
\caption{Widths of the decay modes of the $G_{KK}$ for $M_{G_{KK}}=2.4\units{TeV}$.\label{tab_GKKWidth}}
\end{table}

\subsubsection{$p~\!p\rightarrow \left\{G_{KK},~\!A_{KK},~\!Z_{KK}\right\} \rightarrow t~\!\bar{t} $\label{ttbar}}

\begin{figure}[ht]
\centering
\subfigure[]{
\includegraphics[width=.478\textwidth]{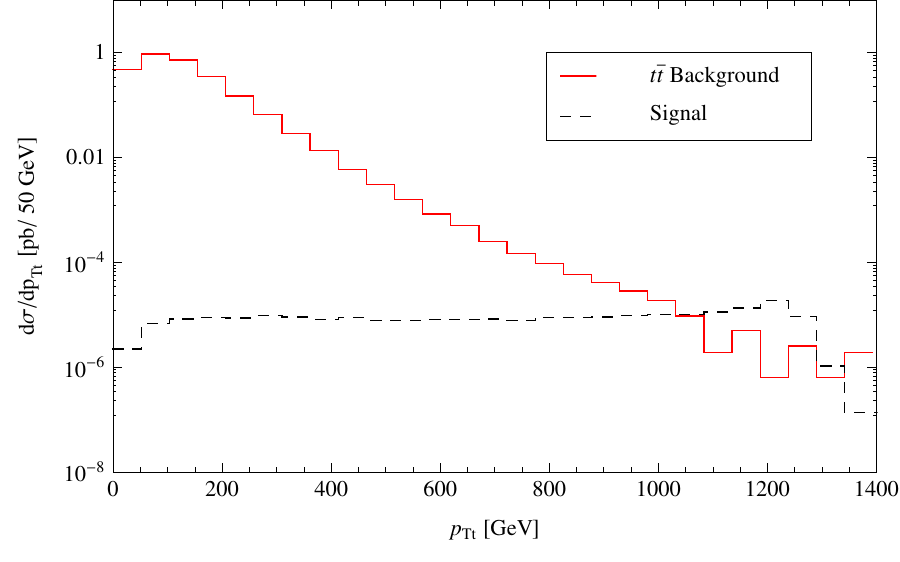}
\label{fig_cutPT8}
}
\subfigure[]{
\includegraphics[width=.478\textwidth]{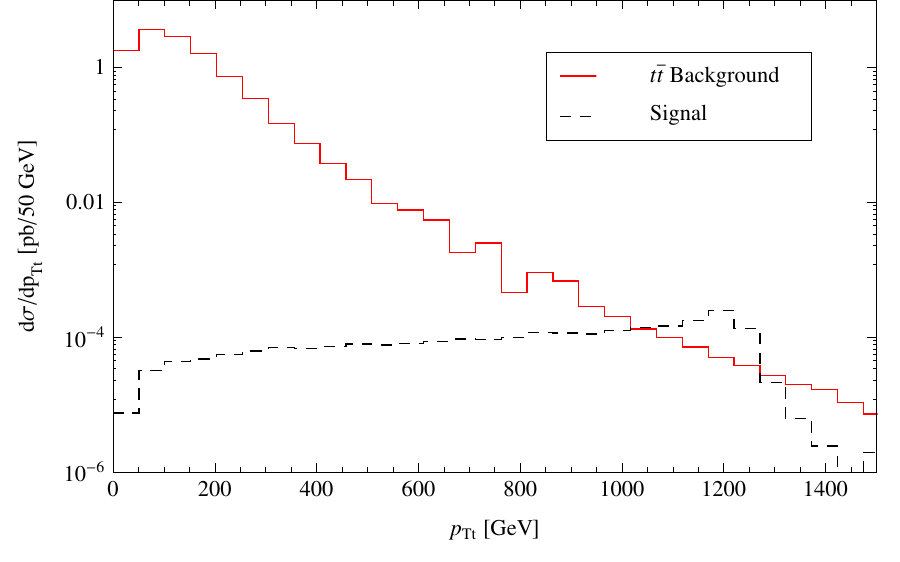}
\label{fig_cutPT14}
}
\caption{(a) Differential cross section from $pp\rightarrow t\bar{t}$ for the transverse momentum of the $t$ at $\sqrt{s} = 8$ TeV. (b) The same for $\sqrt{s} = 14$ TeV}
\end{figure}

The cross sections for producing a $t\overline{t}$ pair through neutral gauge boson KK modes are shown in Table \ref{tab_ttbarproduction}. As stressed above, the production of any hadronic final state is clearly dominated by $G_{KK}$. For $t\overline{t}$ production, the EW KK gauge bosons only contribute approximately 2$\%$ of the cross section.
\begin{table}[h]
\centering
ÊÊÊ\begin{tabular}{cc}
	\\ 
	\ctoprule
	\multicolumn{2}{c}{$\sqrt{s} = 8$ TeV} \\  
ÊÊÊÊÊÊÊ	$X$		&  $\sigma (p p  \rightarrow  X \rightarrow  t\overline{t}) $[pb] \\\cmrule
ÊÊÊÊÊÊÊÊ	$A_{KK}$	& $1.28\cdot10^{-4}$ \\ 
	$Z_{KK}$ 	& $8.79\cdot10^{-5}$ \\ 
	$G_{KK}$	& $1.11\cdot10^{-2}$ \\
	\cmrule
	All 		& $1.13\cdot10^{-2}$ \\ 
	\cbottomrule
	\\
	\ctoprule	
	\multicolumn{2}{c}{$\sqrt{s} = 14$ TeV} \\  
ÊÊÊÊÊÊÊ	$X$		&  $\sigma (p p  \rightarrow  X \rightarrow  t\overline{t}) $[pb]\\ \cmrule
ÊÊÊÊÊÊÊÊ	$A_{KK}$	& $1.68\cdot10^{-3}$\\ 
	$Z_{KK}$ 	& $1.32\cdot10^{-3}$ \\ 
	$G_{KK}$	& $0.118~~$\\
	\cmrule
	All 		&$0.121~~$ \\ 
	\cbottomrule	
ÊÊÊÊ\end{tabular}
\caption{Cross sections for $pp\rightarrow \left\{G_{KK},~\!A_{KK},~\!Z_{KK}\right\} \rightarrow t\overline{t}$ at $\sqrt{s} = 8$ TeV and $\sqrt{s} = 14$ TeV.\label{tab_ttbarproduction}}
\end{table}
The cross sections for producing $t\overline{t}$ are on the order of a few fb, even for the LHC at $\sqrt{s} = 8$ TeV, which is encouraging when looking for a signal. However, the $t\overline{t}$ SM background is pretty large, with a cross section of 139 pb (569 pb) at $\sqrt{s} = 8$ TeV (14 TeV). 
Figures \ref{fig_cutPT8} and \ref{fig_cutPT14} show the transverse momentum distributions of the top quark at 8 and 14 TeV respectively. 

\begin{figure}[ht]
\centering
\subfigure[]{
\includegraphics[width=.478\textwidth]{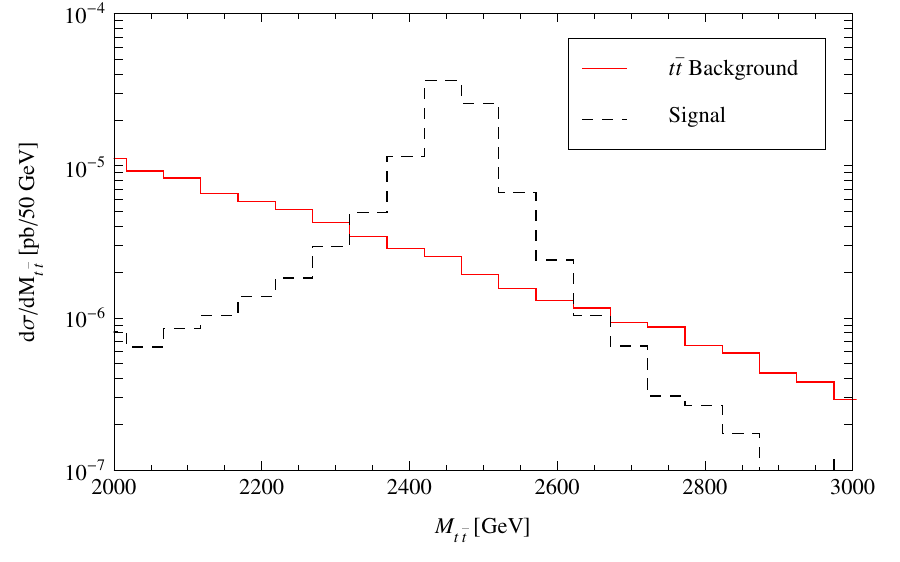}

\label{fig_minvtt8}
}
\subfigure[]{
\includegraphics[width=.478\textwidth]{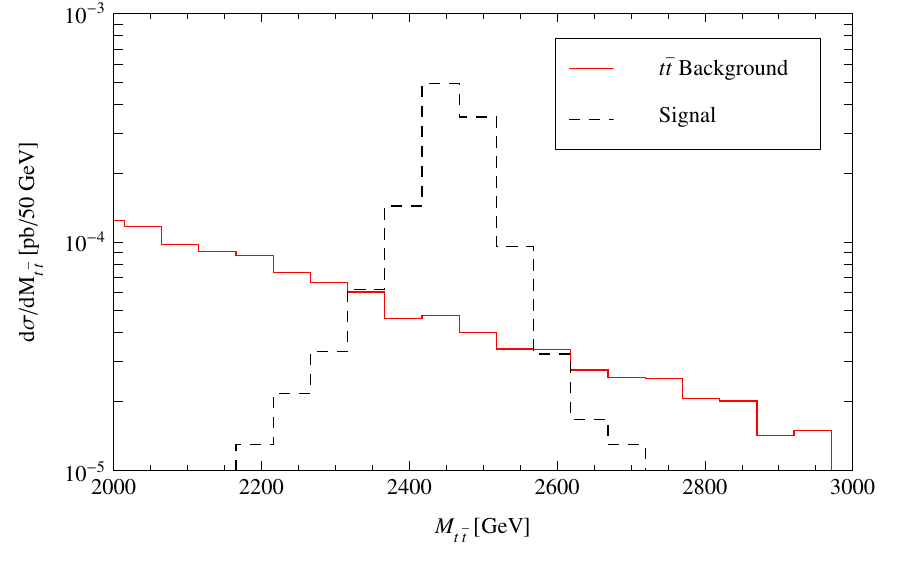}
\label{fig_minvtt14}
}
\caption{(a) Differential cross section as a function of the $t\bar{t}$ invariant mass for $\sqrt{s} = 8$ TeV. (b) The same for $\sqrt{s} = 14$ TeV. }
\end{figure}

At both 8 and 14 TeV we see that a large fraction of the SM cross section is cut out when demanding that
\be
p_{Tt} \ge 800 \text{ GeV}.
\ee
After this cut, the KK mode cross section is 5.11 fb (67.2 fb) with a background of 8.1 fb (103 fb) for  $\sqrt{s}=8$ TeV (14 TeV). The differential cross section for $t\overline{t}$ production is plotted against the invariant mass in Figures \ref{fig_minvtt8} and \ref{fig_minvtt14}. As expected, we observe a clear signal peaked at invariant masses around 2400 GeV. This signal exceeds the background for approximately $2300\units{GeV} \le M_{t\overline{t}} \le 2600\units{GeV}$. 

We thus counted the number of events and computed the significance of the signal for an integrated luminosity of 10 fb$^{-1}$. These are given in Table \ref{tab_8TeVttbar} for a center of mass energy of 8 TeV while Table \ref{tab_14TeVttbar} shows the center of mass energy at 14 TeV. According to the results in Table \ref{tab_8TeVttbar}, with the previous cut the LHC should be able to discover the KK gluons with a mass of 2.4 TeV with 10 fb$^{-1}$ of integrated luminosity at $\sqrt{s} = 8$ TeV.

\begin{table}[h]
\centering
	\begin{tabular}{l | cc |cc}
		\ctoprule
		8 TeV & \multicolumn{2}{ c| }{ Basic } &\multicolumn{2}{ c }{ $M_{t\overline{t}}\units{[TeV]}$ }\\
& \multicolumn{2}{ c| }{ } &\multicolumn{2}{ c }{ $\in[2300,2600]$ }\\
		\cmrule
		10 fb$^{-1}$& $N_{\mt{Events}}$ & $\frac{\mt{S}}{\sqrt{\mt{B}}}$&	 Events &$\frac{\mt{S}}{\sqrt{\mt{B}}}$ \\
		\cmrule
		$G_{KK}$, $A_{KK}$, $Z_{KK}$	& 51	&	5.7	&45	& 15.9\\
		SM	&80	&		&8	& \\
		\cbottomrule
	\end{tabular}
\caption{Events and significance for $t \overline{t}$ production through KK gauge bosons for $\sqrt{s} = 8$ TeV.\label{tab_8TeVttbar}}
\end{table}

\begin{table}[h]
\centering
	\begin{tabular}{ l |cc |cc}
		\ctoprule
		14 TeV & \multicolumn{2}{ c| }{ Basic } &\multicolumn{2}{ c }{ $M_{t\overline{t}}\units{[TeV]}$ }\\
& \multicolumn{2}{ c| }{ } &\multicolumn{2}{ c }{ $\in[2300,2600]$ }\\
		\cmrule
		10 fb$^{-1}$& $N_{\mt{Events}}$ & $\frac{\mt{S}}{\sqrt{\mt{B}}}$&	 Events & $\frac{\mt{S}}{\sqrt{\mt{B}}}$ \\
		\cmrule
		$G_{KK}$, $A_{KK}$, $Z_{KK}$	& 672	&	21.0	&605	& 47.4\\
		SM &1025	&		&163	& \\
		\cbottomrule
	\end{tabular}
\caption{Events and significance for $t \overline{t}$ production through KK gauge bosons for $\sqrt{s} = 14$ TeV.\label{tab_14TeVttbar}}
\end{table}
%

%--------------------------------------------------------------------------------------------------------------------------------------
%  Charged gauge bosons
%--------------------------------------------------------------------------------------------------------------------------------------

\section{Charged Kaluza-Klein gauge bosons}
\label{Prod_Charged}

We now move to the discussion of the production of the charged gauge bosons. The $W_{KK}^\pm$ productions cross-sections for the LHC at $\sqrt{s} = 8$ and $14$ TeV are 0.697 and 10.1 fb, respectively.  We include both the signal from the $W_{KK}^+$ and the $W_{KK}^-$. Table \ref{tab_WKKWidth} shows the widths for the different decay channels. We get a total width of 46 GeV. As in the case of the neutral vectors, the standard search channel in $\ell \nu_\ell$ is not worthy of attention, as it will be strongly suppressed by the small overlap between the $W_{KK}^\pm$ and left-handed lepton profiles. Again, similar considerations apply for the dijet channel. As is apparent, the $t b$ channel is the most promising one and, as we will see, it provides actual chances of observing these particles at the LHC.

\begin{table}[h]
\centering
ÊÊÊÊ\begin{tabular}{l | c c }
	\ctoprule
	Decay		&	Width	&	Branching \\
	channel		&	[GeV]       	&	Ratio \\
	\cmrule
	$t \overline{b}$	&	33.570	                   &	0.732	\\
	$W^+ Z$		&	\phantom{0}5.764	&	0.126	\\
	$W^+ H$		&	\phantom{0}5.679	&	0.124	\\
	$u \overline{d}$&	\phantom{0}0.351	&	0.008	\\
	$c \overline{s}$	&	\phantom{0}0.350	&	0.008	\\
	$e^+ \nu_e$	&	\phantom{0}0.038	&	$8\cdot 10^{-4}$\\
	$\mu^+ \nu_{\mu}$&	\phantom{0}0.038	&	$8\cdot 10^{-4}$\\
	$\tau^+ \nu_{\tau}$&	\phantom{0}0.038	&	$8\cdot 10^{-4}$\\
	$c \overline{d}$&	\phantom{0}0.019	&	$4\cdot 10^{-4}$\\
	$u \overline{s}$&	\phantom{0}0.019	&	$4\cdot 10^{-4}$\\ 
	$c \overline{b}$&	\phantom{0}$6\cdot 10^{-5}$	&	$1\cdot 10^{-6}$\\ 
	$t \overline{s}$&	\phantom{0}$6\cdot 10^{-5}$	&	$1\cdot 10^{-6}$\\ 
	$t \overline{d}$&	\phantom{0}$1\cdot 10^{-5}$	&	$2\cdot 10^{-7}$\\ 
	$u \overline{b}$&	\phantom{0}$4\cdot 10^{-6}$	&	$8\cdot 10^{-8}$\\ 
	\cmrule
	Total			&	45.866	&		\\ 
	\cbottomrule
	\end{tabular}
\caption{Widths of the decay modes of the $W_{KK}^+$  for $M_{W^\pm_{KK}}=2.4\units{TeV}$.\label{tab_WKKWidth}}
\end{table}

\subsection{$p~\!p\rightarrow W_{KK} \rightarrow t~\!b$}

Let us then focus on the signal from the $W_{KK}$ decaying into top and bottom quarks, followed by a semileptonic decay of the top, $t\rightarrow W b\rightarrow \ell \nu_\ell b$. The cross section for producing the $W_{KK}$ were calculated for $\sqrt{s} = 8$ and $14$ TeV. These are shown in Table \ref{tab_wkk_bblvl}. For completeness, we also show in that table the computation for the $W^+Z$ channel, as we did in the case of the neutral KK gauge bosons. As in that case, however, the small branching ratio for the decay of $W_{KK} \rightarrow WZ$ makes this signal not worthy of consideration even for $\sqrt{s}=14\units{TeV}$.

\begin{table}[h]
\centering
ÊÊÊÊ\begin{tabular}{ccccc}
	\\ 
	\ctoprule
ÊÊÊÊÊÊÊ	$\sqrt{s}$	&  $\sigma (p p \!\rightarrow\!  W_{KK}\rightarrow\! t b) $[pb] 	&BR$(t\!  \rightarrow\!  W b)$	&BR$(W\! \rightarrow\! \ell \nu_{\ell})$&	Total [pb]\\ 
	\cmrule
	8		& $5.10\cdot10^{-4}$ 							& 1 						& 0.216					&	$1.10\cdot10^{-4}$ \\
	14 		& $7.36\cdot10^{-3}$ 							& 1 						& 0.216					&	$1.59\cdot10^{-3}$ \\ 
	\cbottomrule\\
	\ctoprule
	$\sqrt{s}$	&  $\sigma (p p\!  \rightarrow\!  W_{KK}\rightarrow\! W Z) $[pb] 	& BR$(W\!  \rightarrow\!  \ell \nu_\ell)$	& BR$(Z\! \rightarrow\! \ell^+ \ell^-)$	&	\!\!Total [pb]\\ 
	\cmrule
	8		& $ 8.76\cdot10^{-5}$							& 0.216					& 0.067					&	$1.27\cdot10^{-6}$ \\
	14		& $ 1.26\cdot10^{-3}$							& 0.216					& 0.067					&	$1.84\cdot10^{-5}$ \\ 
	\cbottomrule		
ÊÊÊÊ\end{tabular}
\caption{Cross sections for $p p \rightarrow W_{KK} \rightarrow b \overline{b} \ell \nu_{\ell}$ and $p p \rightarrow W_{KK}\rightarrow \ell \nu_\ell \ell \ell$.\label{tab_wkk_bblvl}}
\end{table}

\begin{figure}[ht]
\centering
\subfigure[]{
\includegraphics[width=.478\textwidth]{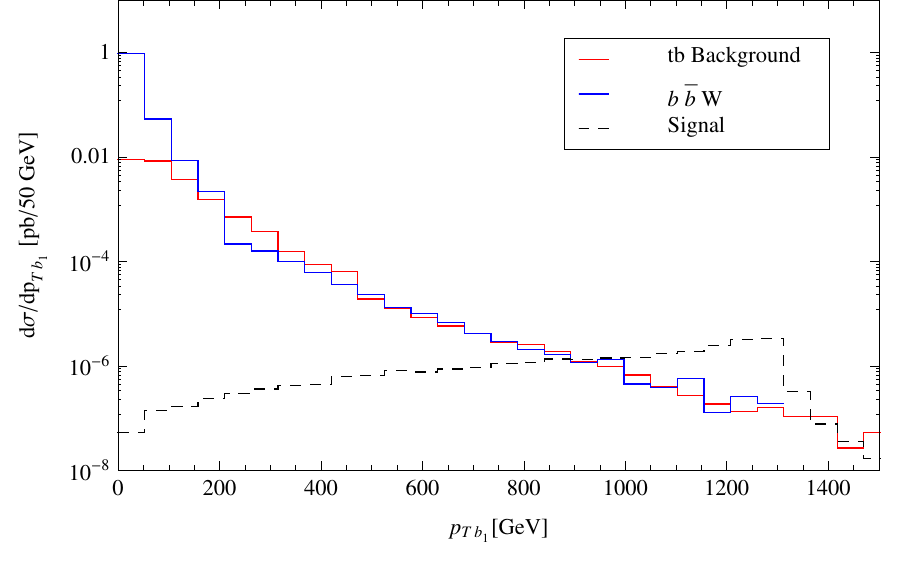}
\label{fig_cutPTb1}
}
\subfigure[]{
\includegraphics[width=.478\textwidth]{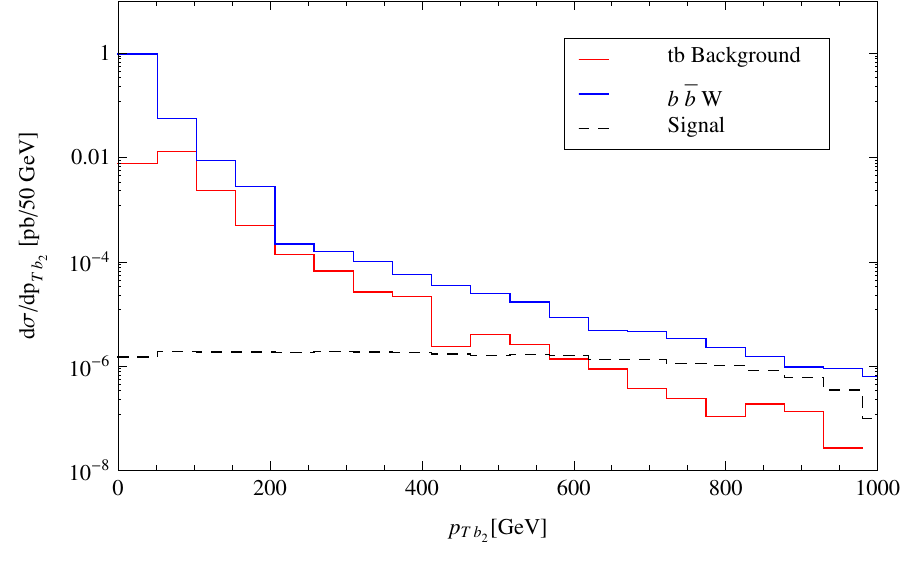}
\label{fig_cutPTb2}
}
\subfigure[]{
\includegraphics[width=.478\textwidth]{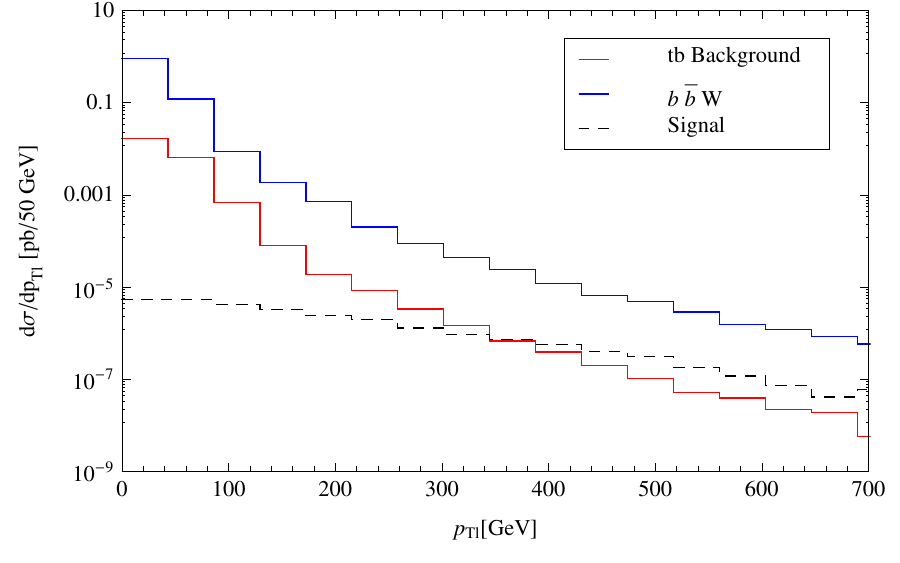}
\label{fig_cutptl}
}
\subfigure[]{
\includegraphics[width=.478\textwidth]{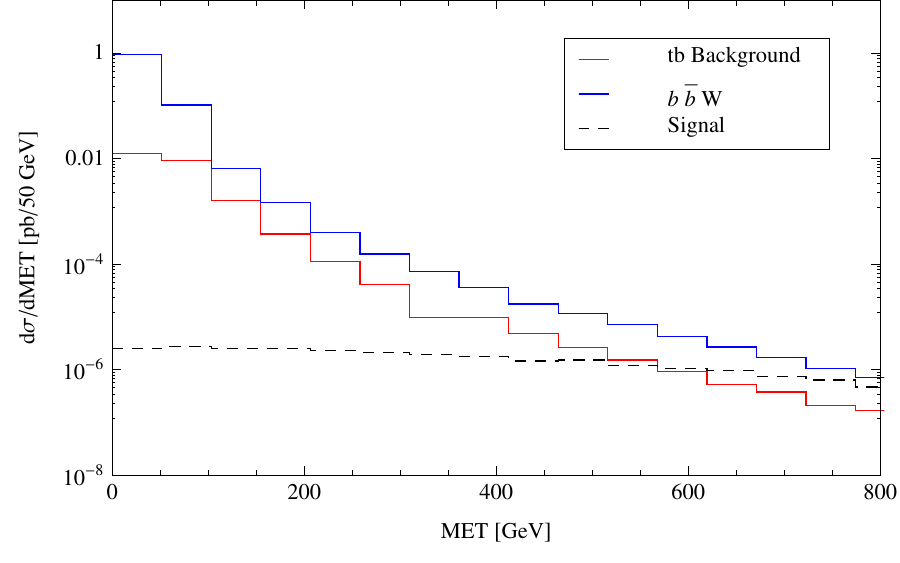}
\label{fig_cutmet}
}
\subfigure[]{
\includegraphics[width=.478\textwidth]{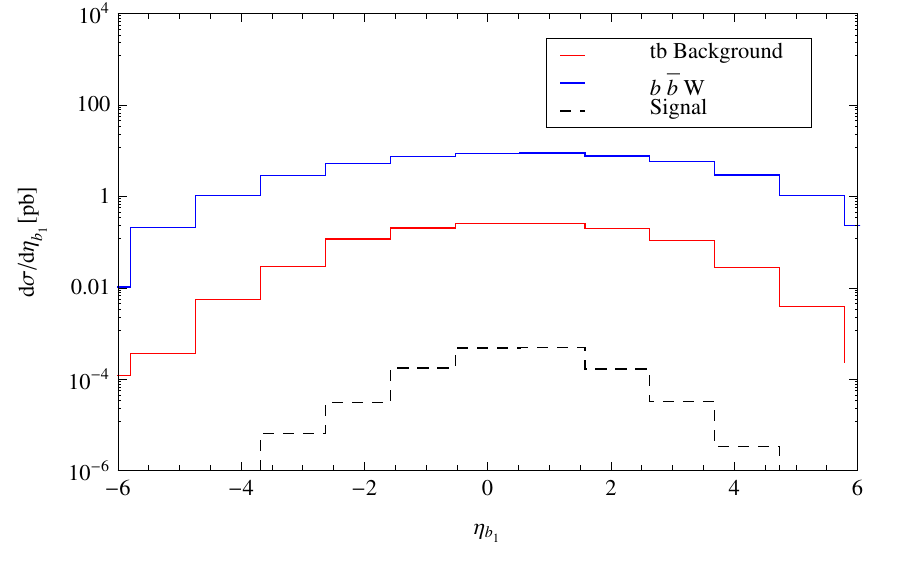}
\label{fig_yb1}
}
\subfigure[]{
\includegraphics[width=.478\textwidth]{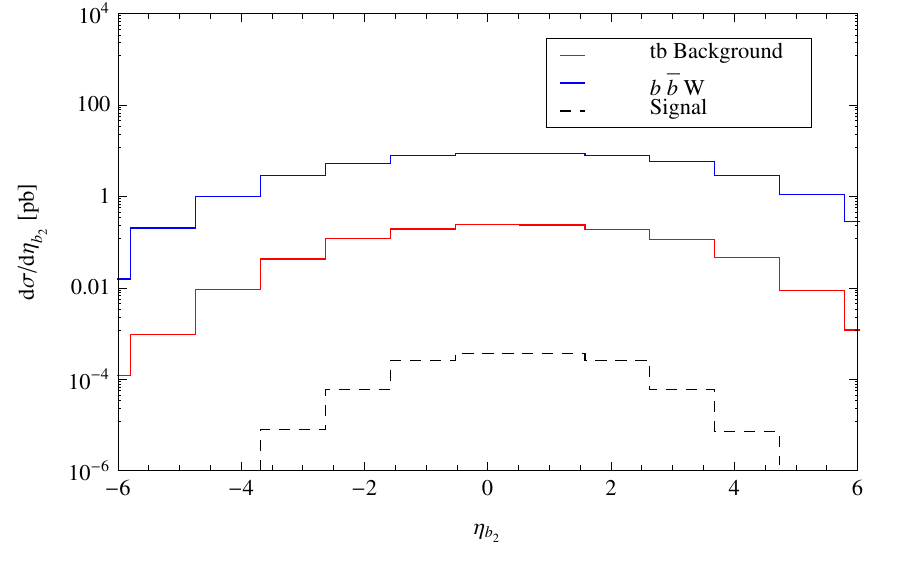}
\label{fig_yb2}
}
\caption[]{Differential cross sections from $pp \rightarrow  t\bar{b}$ for \subref{fig_cutPTb1} the transverse momentum of the primary $b$ quark, \subref{fig_cutPTb2} transverse momentum of the $b$ quark coming from the $t$ decay, \subref{fig_cutptl} transverse momentum of the charged lepton, \subref{fig_cutmet} missing transverse energy, \subref{fig_yb1} rapidity of the primary $b$ quark, and \subref{fig_yb2} rapidity of the $b$ quark coming from the $t$ decay.}
\label{fig_CutsWtb}
\end{figure}

\begin{figure}[ht]
\centering
\subfigure[]{
\includegraphics[width=.478 \textwidth]{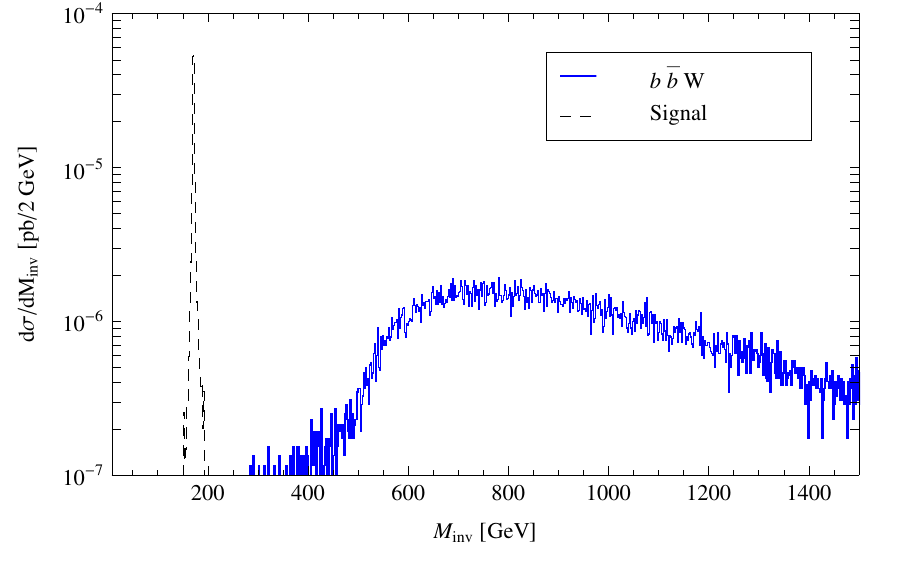}
\label{fig_blvl_InvMass}
}
\subfigure[]{
\includegraphics[width=.478 \textwidth]{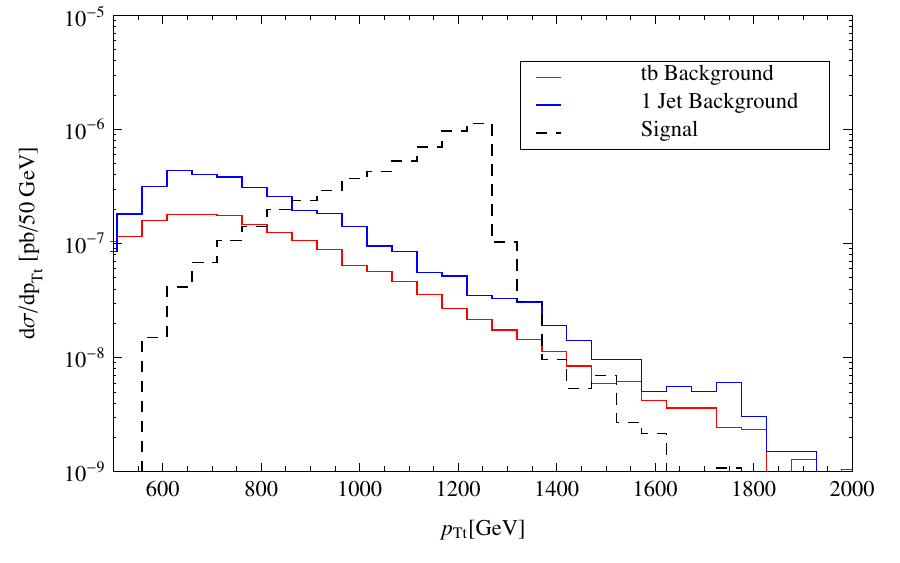}
\label{fig_tb_PTt}
}
\subfigure[]{
\includegraphics[width=.478 \textwidth]{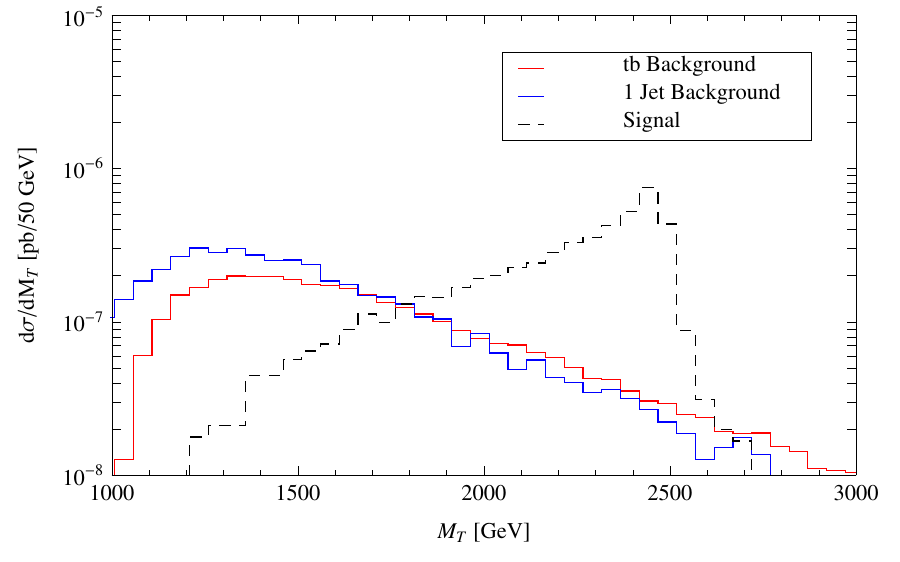}
\label{fig_tb_MT}
}
\subfigure[]{
\includegraphics[width=.478 \textwidth]{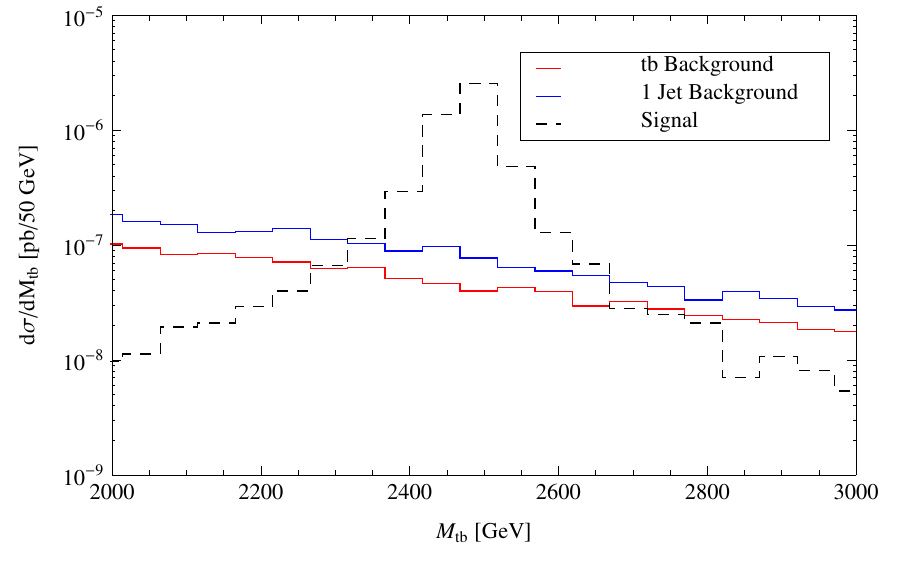}
\label{fig_tb_Minv}
}
\caption[fig_WtbCuts]{Differential cross section for $pp \rightarrow b\overline{b}\ell\nu_\ell$. In all panels the black dashed line denotes the KK mode signal, while solid lines are used for the different backgrounds. Panel \subref{fig_blvl_InvMass} shows the $b\overline{b}W$ background can be neglected by demanding the reconstructed top mass be at the mass of the top. \subref{fig_tb_PTt} Differential cross section as a function of the transverse momentum of the reconstructed top. \subref{fig_tb_MT} The same for the total transverse mass. \subref{fig_tb_Minv} The same for the total invariant mass.}
\end{figure}

We thus focus on the $t b$ signal at 14 TeV. The irreducible backgrounds for this process are $W\rightarrow t b$ and $p p \rightarrow g W \rightarrow b\overline{b}\ell\nu_\ell$. We will refer to the latter as the $b\overline{b}W$ background. This process does not go through a top quark, so reconstructing the top mass from the $b\ell\nu_\ell$ should significantly reduce this background. On the other hand, we expect the KK decay to produce final state $b$'s with large transverse momentum. Shown below are the differential cross sections generated with MadGraph 5 for the transverse momentum of the initial $b$ in Figure \ref{fig_cutPTb1} and the second $b$ in Figure \ref{fig_cutPTb2}. Similarly, the transverse momentum of the lepton and the missing energy should be greater than in the SM. These shown in Figures \ref{fig_cutptl} and \ref{fig_cutmet} respectively.
We also look at the rapidity of the $b$-jets. These are shown in Figure \ref{fig_yb1} for the first $b$ and Figure \ref{fig_yb2} for the second $b$. From these, we infer the following cuts to decrease the SM background without decreasing the total cross section significantly:
\begin{eqnarray}
p_{Tb}	&\ge&	200 \text{ GeV}, \nonumber \\
p_{T\ell} &\ge&	150 \text{ GeV}, \nonumber \\
\cancel{E}_T	&\ge&	150 \text{ GeV}, \nonumber \\
\left| \eta_{b} \right|	&\le&	3.0.
\label{Wtbcuts}
\end{eqnarray}
After making these cuts, the KK mode cross section is 0.27 fb. The SM background cross section is 0.17 fb while the $b\overline{b}W$ cross section is 1.16 fb. In Figure \ref{fig_blvl_InvMass} the differential cross section is shown for the reconstructed top mass. The actual events fall near the top mass while the $b\overline{b}W$ background is very spread. Demanding that the reconstructed top mass falls between 150 and 200 GeV reduces the  $b\overline{b}W$ cross section to $1.89\cdot10^{-3}$ fb, which is two orders of magnitude lower than the KK signal. Assuming this extra cut, we can ignore this background for the rest of the analysis. 

There are other important backgrounds. In particular, one important source of background comes from the gluon-$W$ fusion process with an additional jet. Before making any cuts, the cross section for $p p \rightarrow tbj\rightarrow b\overline{b} \ell \nu_\ell$ is 27.4 pb. This is reduced to 0.834 fb after applying all the cuts in (\ref{Wtbcuts}). Still, this is larger than the signal but can be further suppressed by requiring again a cut in the $p_T$ of the extra jet. For instance, using the same hard cut as for the $b$-jets, $p_{Tj} \ge 200 \units{GeV}$, in the events with an extra jet, the cross sections goes to 0.255 fb, below the signal. Finally, reducing other large backgrounds like $W$ production in association with light jets misidentified as $b$-jets would require, in addition, the use of $b$-tagging techniques.

We have searched for the signal in three different variables, the reconstructed transverse momentum of the top quark, the reconstructed total transverse mass, $M_T$, and the total invariant mass, $M_{tb}$. Again, we counted the number of events for integrated luminosities which might permit unveiling the presence of the new charged gauge bosons. In this case 10 fb$^{-1}$ proves to be too small for a discovery. We calculated the number of events and the significance of the signal after implementing the basic cuts for 150 fb$^{-1}$. Finally, we also focus in the different regions of interest in the three variables considered. For instance, from Figure \ref{fig_tb_PTt} we observe that signal exceed the background for transverse momentum of the reconstructed top between $\sim800$ and 1400 GeV. Figure \ref{fig_tb_MT} motivates us to look at the signal for the total transverse mass in the range of 1800 and 2500 GeV. The best variable for getting the mass of the $W_{KK}$ is the total invariant mass. In this case the excess is located between 2400 and 2700 GeV. The results are summarized in Table \ref{tab_150fb_wkktb}. There we also show the number of events resulting from demanding one or two $b$ tags in the events. We assume a naive $b$-tagging efficiency of $\epsilon_b=0.6$.

\begin{table}[h]
\centering
	\begin{tabular}{l | cc | cc | cc | cc}
		\ctoprule
		\!\!\!14 TeV\!\!\!& \multicolumn{2}{ c| }{ Basic } &\multicolumn{2}{ c| }{ $p_{Tt}\units{[GeV]}$ }& \multicolumn{2}{ c| }{ $M_T\units{[GeV]}$} &\multicolumn{2}{ c}{$M_{tb}\units{[GeV]}$} \\
 & \multicolumn{2}{ c| }{ } &\multicolumn{2}{ c| }{ $\in[800, 1400]$ }& \multicolumn{2}{ c| }{ $\in[1800,2500]$} &\multicolumn{2}{ c}{$\in[2400,2700]$}\!\! \\
		\cmrule
		\!\!\!150 fb$^{-1}$\!\!& $N_{\mt{Events}}$ & $\frac{\mt{S}}{\sqrt{\mt{B}}}$&$N_{\mt{Events}}$& $\frac{\mt{S}}{\sqrt{\mt{B}}}$&$N_{\mt{Events}}$& $\frac{\mt{S}}{\sqrt{\mt{B}}}$&$N_{\mt{Events}}$& $\frac{\mt{S}}{\sqrt{\mt{B}}}$ \\
		\hline
		\!\!\!$W_{KK}$\!\!\!\!	&40&5.0 &38&9.5&34&9.4&36&16.1\\
			                   &\!\!(24)~\![14]\!&\!(3.9)~\![3.0]\!\!&\!\!(23)~\![13]\!&\!(7.7)~\![5.3]\!\!&\!\!(20)~\![12]\!&\!(7.1)~\![6]\!\!&\!\!(22)~\![13]\!&\!(12.7)~\![13]\!\! \\
		&&&&&&&&\\
		\!\!\!SM\!\!\!\!              &63&&16&&13&&5&\\
		                             &\!\!(37)~\![22]\!&&\!\!(9)~\![6]\!&&\!\!(8)~\![4]\! &&\!\!(3)~\![1]\! &\\
		\cbottomrule
	\end{tabular}
\caption{Events and significance for $W_{KK}\rightarrow b\overline{b}\ell\nu_\ell$. In parentheses (backets) we give the corresponding numbers imposing one (two) $b$ tagging(s). \label{tab_150fb_wkktb}}
\end{table}

The results of Table \ref{tab_150fb_wkktb} show that the LHC should be able to discover the first KK mode of the $W^{\pm}$ gauge boson with a mass of 2.4 TeV, with 150 fb$^{-1}$ of integrated luminosity.  \footnote{An important caveat must be stressed here. These numbers have been obtained not taking into account $t\overline{t}$ production as a source of background. Indeed a boosted top decaying hadronically can fake a $b$-jet. Moreover, this background comes not only from the SM but also from the $s$-channel exchange of the KK gluons, whose mass is the nearly the same as the $W_{KK}^\pm$. An effective discrimination between tops and $b$-jets would be possible by imposing additional cut in the jet-mass variable. Following the analysis in \cite{Agashe:2008jb}, an appropriate cut results in an acceptance of $\sim 50\%$ of the $b$-jets, while rejects almost all the $t$-jets. This kind of analysis goes beyond the parton-level analyses employed in this paper. At any rate, extrapolating the previous result to our case we still have enough significance for a discovery with $L=150\units{fb}^{-1}$.}

%-------------------------------- DOCUMENT: CONCLUSIONS----------------------------------%

\section{Conclusions}

In this paper we have discussed the possible signals at the LHC for a recent class of warped extra-dimensional scenarios, with a bulk Higgs and a deformation of the $AdS_5$ metric near the IR brane \cite{Cabrer:2011fb}. This modification can soften the constraints from EWPD, allowing for relatively light KK excitations. Moreover, this can be done within a minimal extension of the SM field content. It only requires an extra bulk scalar field, the radion, in order to stabilize the size of the extra dimension. Thus, this kind of models provide an appealing alternative to those scenarios with gauge custodial symmetry in the bulk, and hence a more complicated spectrum.

We have studied the most common signals where one expects to observe extra neutral and charged resonances in extra dimensional models. As we show, only signals involving top and bottom decay channels offer good chances to probe this kind of models at the LHC. The reason, which also explains why we can alleviate current bounds from EW precision observables, is that in this kind of models one of the main features is a stronger localization of KK gauge bosons towards the IR than in the standard $AdS_5$ scenarios, while the Higgs is pushed into the bulk. This results in suppressed couplings to the SM particles other than those to third family of quarks. 

We have discussed the standard leptonic and diboson channels where no signal is expected for the LHC energies and realistic luminosities. We then focus on signals involving the third generation and study $t\overline{t}$ production, which is dominated by the KK gluon exchange, as well as $t b$ in the case of the charged KK gauge bosons. After making the adequate cuts, the LHC should be able to probe the existence of the KK gluons for $\sqrt{s}=8\units{TeV}$ and 10 fb$^{-1}$. Testing the charged EW KK gauge bosons, however, would require waiting until $\sqrt{s}=14\units{TeV}$ and larger integrated luminosities. We emphasize again that in this kind of models these are the only distinct signal expected from new physics. Therefore, should the existence of the KK gauge excitations be probed, one could distinguish this scenario from other models like those with custodial symmetry, where not only signals in the diboson channel are also expected \cite{Agashe:2007ki,Agashe:2008jb}, but in general predict relatively light new fermions.

%------------------------------------ ACKNOWLEDGEMENTS ---------------------------------------%
\section*{Acknowledgements}
It is a pleasure to thank M. Quir\'os for useful discussions. This work has been supported in part by the U.S. National Science Foundation under Grant PHY-0905283-ARRA.

%\newpage

%------------------------------------------- REFERENCES -------------------------------------------%

\bibliographystyle{apsrev}

%-------------------------------------------------- END --------------------------------------------------%

\end{document}